\documentclass[5p,twocolumn,sort&compress]{elsarticle}
\usepackage[thicklines]{cancel}
\usepackage{graphicx}      
\usepackage{amsmath,amsfonts,amssymb}  
\usepackage{hyperref}      
\usepackage{bm}            
\usepackage{subfigure}     
\usepackage{color}         
\usepackage{xspace}

\usepackage{ulem}
\newcommand{\snn} {\ensuremath{\sqrt{s_{\rm NN}}}\xspace}
\newcommand{\pt} {\ensuremath{p_{\rm T}}\xspace}

\usepackage{comment}
\begin{document}

\begin{frontmatter}

 \title{Tracing \pt-differential radial flow from blast-wave analytics to quark coalescence}

\author[label1,label2]{Jie Wan\corref{cor1}}
\ead{21110200012@m.fudan.edu.cn}
\author[label1,label2]{Chun-Zheng Wang}
\ead{wangcz22@m.fudan.edu.cn}
\author[label1,label2]{Yu-Gang Ma\corref{cor1}}
\ead{mayugang@fudan.edu.cn}
\author[label1,label2]{Qi-Ye Shou\corref{cor1}}
\ead{shouqiye@fudan.edu.cn}
\cortext[cor1]{Corresponding authors}
\address[label1]{Key Laboratory of Nuclear Physics and Ion-beam Application (MOE), Institute of Modern Physics, Fudan University, Shanghai 200433, China}
\address[label2]{Shanghai Research Center for Theoretical Nuclear Physics, NSFC and Fudan University, Shanghai 200438, China}

\begin{abstract}
The observable $v_0(\pt)$, which quantifies event-by-event fluctuations in the differential transverse-momentum spectrum, is proposed as a direct and penetrating probe of radial flow in heavy-ion collisions.
Recent measurements at the LHC exhibit a clear mass ordering for pions, kaons and protons at low \pt and a baryon-meson splitting at intermediate \pt, resembling to the well-known features of elliptic flow ($v_2$).
In this letter, we first derive an analytic expression of $v_0(\pt)$ within a Blast-Wave framework incorporating fluctuations of freeze-out temperature and radial expansion velocity, which can naturally explains the experimentally observed mass ordering.
The distinct dynamical origins of the mass ordering in $v_0(\pt)$ and $v_2(\pt)$ are discussed.
Furthermore, using the AMPT model, we demonstrate that the baryon-meson splitting emerges spontaneously from the quark coalescence. This study provides deeper insight into the $v_0(\pt)$ observable and the collective dynamics of the QGP.
\end{abstract}

\end{frontmatter}

\section{Introduction}

Heavy-ion collisions at the LHC and RHIC create temperatures and energy densities well above the QCD crossover, leading to the deconfinement of quarks and gluons into a short-lived quark–gluon plasma (QGP)~\cite{alice_qgp, nst_alice, nst_star}. This medium undergoes collective expansion driven by large pressure gradients, which contains two distinct components~\cite{flow_review1}: an anisotropic expansion that converts initial spatial asymmetries into final-state momentum anisotropies, and an isotropic (radial) expansion corresponding to the collective transverse boost of all produced particles.

The azimuthal anisotropy of particle emission is conventionally characterized by a Fourier expansion~\cite{Voloshin:1996, flow_ep}.
The elliptic coefficient $v_{2}$ provides the first clear evidence of the near-ideal fluid behavior of the QGP, establishing a stringent upper bound on the specific shear viscosity $\eta/s$ ~\cite{v2_fluid}.
Key observations including mass ordering, baryon-meson splitting, number of constituent quark (NCQ) scaling, as well as higher-order harmonics extensively reveal the hydrodynamic and transport properties of the QGP~\cite{flow_review1}. 

In contrast, the study of isotropic expansion is less explored. It traditionally relies on blast-wave (BW) or hydrodynamic fits to identified hadron spectra~\cite{STAR:2004,ALICE:2020,Retiere:2004}. These approaches extract only a single average radial velocity per centrality class, thereby obscuring any possible dependence on transverse momentum ($\pt$) and masking event-by-event fluctuations. Recently, a new \pt--differential observable was proposed in Ref.~\cite{shen_v0}:
\begin{equation}
  v_{0}(\pt) \equiv
  \frac{\langle \delta N(\pt)\, \delta \pt \rangle}
       {\langle N(\pt) \rangle\, \sigma_{\pt}},
  \qquad
  \sigma_{\pt}^{2} = \langle (\delta \pt)^{2} \rangle,
  \label{eq:v0_original}
\end{equation}
where $\delta N(\pt)$ denotes the fluctuation of the yield in a given $\pt$ bin relative to its ensemble mean, $\delta \pt = [\pt] - \langle [\pt] \rangle$ is the fluctuation of the event-wise mean transverse momentum $[\pt]$ at fixed multiplicity, and $\sigma_{\pt}$ is the standard deviation of $[\pt]$ across events. 
Such an observable quantifies the correlation between event-by-event spectral shape fluctuations and the global radial boost, providing direct sensitivity to fluctuations in collective radial expansion and their microscopic origin.

The first experimental measurement of $v_{0}(\pt)$ was reported by the ATLAS Collaboration~\cite{ATLAS_v0}, demonstrating its sensitivity to the bulk viscosity of the QGP. 
The ALICE Collaboration extended the study to pions ($\pi$), kaons ($K$), and protons ($p$)~\cite{ALICE_v0}, revealing clear mass ordering at low $\pt$ and a pronounced baryon-meson splitting at intermediate $\pt$, resembling the well-known features of $v_2$. 
These intriguing results have attracted considerable interest~\cite{v0_2024, Swati_bw, Parida2025, Bhatta2025, STAR2025, Jia2025, Du2025}.
Hydrodynamic simulations~\cite{v0_2024}, which attribute $v_{0}(\pt)$ primarily to temperature or entropy-density fluctuations on the freeze-out surface, successfully capture the rising trend with $\pt$ and the mass ordering. A recent BW+Bayesian analysis~\cite{Swati_bw} further supports such a species-dependent behavior.
Nevertheless, previous simulation studies have neither provided a clear analytical form for $v_{0}(\pt)$ nor systematically investigated the origin of the baryon-meson splitting, both of which are essential pieces for a complete understanding of this novel observable.

To address these questions, we present in this letter an analytic parametrisation of $v_{0}(\pt)$ derived within the BW framework incorporating event-by-event fluctuations of the kinetic freeze-out temperature and the radial flow velocity, and show how it gives rise to the observed mass ordering. 
Furthermore, using A Multi-Phase Transport (AMPT) model which includes quark transport and coalescence, we discuss the microscopic origin of the baryon-meson splitting of $v_{0}(\pt)$. 

\section{$v_0(\pt)$ from thermal freeze-out dynamics}
\subsection{Parametrization of $v_0(\pt)$ in the blast-wave model}

The blast-wave (BW) model based on the thermal phenomenology is widely used in the study of heavy-ion collision to quantitatively describe the transverse momentum (\pt) spectra and also azimuthal anisotropies in the low-\pt region.
The model assumes that hadrons are emitted from a locally thermalised, collectively expanding fireball at kinetic freeze-out. In the transverse plane, the position of a fluid element is parametrised as
\(
x = R_{x} \hat{r} \cos(\hat{\phi}), \quad y = R_{y} \hat{r} \sin(\hat{\phi}),
\)
where $R_x$ and $R_y$ are the spatial radii of the emission source along the $x$- and $y$-axes, respectively, and $\hat{r} \in [0,1]$, $\hat{\phi} \in [0,2\pi]$ are the normalized radial and azimuthal coordinates. The collective expansion is characterized by a transverse flow rapidity profile,
$\rho = \hat{r} (\rho_0 + \rho_2 \cos(2\phi_b))$,
where $\tan \phi_b = R_x/R_y \tan \hat{\phi}$ and  $\quad m_T = \sqrt{p_T^2 + m^2}$.
Here, $\rho_0$ describes the isotropic radial expansion, and $\rho_2$ introduces an 2-order azimuthal anisotropy, corresponding to the elliptic flow coefficient $v_2$. 

The invariant yield of particles is given by~\cite{Retiere:2004, BW_1993}
\begin{equation}
\frac{d^{2}N}{2\pi \pt\,d\pt\,dy}
= A \!
  \int_{0}^{1}\!\hat r\,d\hat r
  \int_{0}^{2\pi}\!d\hat\phi\;
  m_{T}\,
  I_{0}(\alpha)\,
  K_{1}(\beta),
\label{eq:bw_yield}
\end{equation}
where $T$ is the kinetic freeze-out temperature, and $I_{n}$ and $K_{n}$ are modified Bessel functions. The terms $\alpha = \pt\sinh\rho/T$ and $\beta  = m_{T}\cosh\rho/T$ are defined
with $\rho(\hat r,\hat\phi)$ being the transverse flow rapidity and $m_{T}=\sqrt{\pt^{2}+m^{2}}$ the transverse mass.

We perform a simultaneous fit to $\pi$, $K$, and $p$ \pt spectra and elliptic flow data measured by ALICE~\cite{ALICE:2020,ALICE_v2} in Pb--Pb collisions at $\snn = 5.02$\,TeV for 30--40\% centrality. The radius $R_y$ is fixed at 10.0\,fm. Extracted model parameters are summarized in Tab.~\ref{tab:params}, compared with previously reported values~\cite{ALICE:2020} from ALICE. The parameter $\beta_s = \tanh(\rho_0)$ represents the surface velocity. The fitted freeze-out temperature shows good agreement with published results.

\begin{table}[htbp]
  \centering
  \caption{Blast Wave fit parameters for 30–40\% Pb–Pb collisions at \snn = 5.02 TeV.}
  \label{tab:params}
  \begin{tabular}{lcc}
    \hline
    Parameters           & This work & ALICE~\cite{ALICE:2020} \\
    \hline
    \(T\) (GeV)              & \(0.108 \pm 0.002\)       & \(0.101\pm 0.003\)      \\
     \(\beta_{s}\) & \(0.837 \pm 0.002\)  & \(0.622 \pm 0.003\)   \\
     \(\rho_{2}\)   & \(0.085 \pm 0.002\) & \(N/A \) \\
     \(R_{x}\)   & \(8.30 \pm 0.013\) & \(N/A \) \\
    \hline
  \end{tabular}
\end{table}

To reproduce the experimentally observed $v_{0}(\pt)$, which reflects event-by-event fluctuations of the transverse-momentum distribution, the bw model here is specifically extended to include independent variations $\delta T$ and $\delta\beta$ around the mean kinetic freeze-out temperature $\bar{T}$ and the surface flow velocity $\bar{\beta}_{s}$.  
The raw yield $N(\pt)$ is replaced by the normalised spectrum $f(\pt)$ to suppress the influence of multiplicity fluctuations,
\begin{equation}
  v_{0}(\pt) =
  \frac{\bigl\langle \delta f(\pt)\,\delta \pt\bigr\rangle}
       {f(\pt)\,\sigma_{\pt}}, \\[4pt]
\label{eq:v0def_alice}
\end{equation}
with \(f(p_T)=\dfrac{N(p_T)}{\int N(p_T)\,\mathrm{d}p_T}\), $\sigma_{\pt}^{2}=\langle(\delta \pt)^{2}\rangle$ and $\delta f(\pt) = f(\pt) - \langle f(\pt)\rangle$.

The resulting modification to the \(p_T\) spectra can be obtained by performing a Taylor expansion of Eq.~(\ref{eq:bw_yield}),
\begin{align}
  \delta N(\pt;T,\beta_{s}) 
  &\approx 
  \left. N'_{T}(\pt) \right|_{\substack{T=\bar{T} \\ \beta_{s}=\bar{\beta}_{s}}} \delta T \notag\\
  &\quad + \left. N'_{\beta_s}(\pt) \right|_{\substack{T=\bar{T} \\ \beta_{s}=\bar{\beta}_{s}}} \delta \beta.
\end{align}
To simplify the notation, the subscripts indicating evaluation of the derivatives at  $T=\bar{T}$ and $beta_{s}=\bar{\beta}_{s}$ will be omitted in the following expressions.
Based on Eq.~(\ref{eq:v0def_alice}), the fluctuations of \(f({\pt)}\) is given by
\begin{align}
\delta f(p_T) \approx \frac{\delta N(p_T)}{N} - \frac{ N(p_T)\delta N}{N^{2}},
\end{align}
where $N$ is the mean multiplicity and \(\delta N\) is the event-wise fluctuation. Assuming that $\delta N/N \ll 1$, the fluctuation \(\delta N\) follows 
\begin{align}
\delta N 
&= \int \!\mathrm{d}p_T\,N'_{T}(\pt)\,\delta T 
   + \int \!\mathrm{d}p_T\,N'_{\beta}(\pt)\,\delta \beta \notag \\
&= M_{T}\,\delta T + M_{B}\,\delta\beta,
\end{align}
where 
\begin{align}
M_{T} =\int \!\mathrm{d}p_T\,N'_{T}(\pt), 
\quad
M_{B} =\int \!\mathrm{d}p_T\,N'_{\beta}(\pt).
\end{align}
For $\delta p_\mathrm{T}$, contributions from all particle species should be accounted for. This leads to:
\begin{align}
\delta p_T
=\int \!\mathrm{d}p_T\,
p_T\,\delta f_{tot}(p_T)
= A\,\delta T + B\,\delta\beta,
\end{align}
where we introduce the shorthand
\begin{align}
A \equiv\; &
\frac{1}{\sum_{i=1}^{n} N_i}
\sum_{i=1}^{n} \int_{0}^{\infty} N'_{i,T}(p_T)\, p_T\, dp_T \notag\\[0.6em]
&\;-\;
\frac{
\sum_{i=1}^{n} M_{i,T}
}{
\left( \sum_{i=1}^{n} N_i \right)^2
}
\sum_{i=1}^{n} \int_{0}^{\infty} N_i(p_T)\, p_T\, dp_T, 
\\[1em]
B \equiv\; &
\frac{1}{\sum_{i=1}^{n} N_i}
\sum_{i=1}^{n} \int_{0}^{\infty} N'_{i,\beta}(p_T)\, p_T\, dp_T \notag\\[0.6em]
&\;-\;
\frac{
\sum_{i=1}^{n} M_{i,B}
}{
\left( \sum_{i=1}^{n} N_i \right)^2
}
\sum_{i=1}^{n} \int_{0}^{\infty} N_i(p_T)\, p_T\, dp_T.
\end{align}

Within the BW framework where all particle species (\(\pi\), \(K\), \(p\)) share the same radial expansion velocity field,  
the event-by-event transverse momentum fluctuation \(\delta \pt\) and its variance \(\sigma_{\pt}^{2}\)  
should be evaluated using \textit{all} particles. This approach is consistent with experimental analyses and ensures that differences in \(v_{0}(\pt)\) across species arise unambiguously from their spectral responses. 

Under the assumptions that \(\langle\delta T\rangle=\langle\delta\beta\rangle=0\) and \(\langle\delta T \delta\beta\rangle\ =0\),
\begin{equation}
\begin{aligned}
\bigl\langle \delta f_{i}(p_T)\,\delta p_T \bigr\rangle
= \; &\frac{1}{N_i} \biggl\langle 
  N'_{i,T}(p_T)\, \delta T 
  + N'_{i,\beta}(p_T)\, \delta\beta \\
&\quad - f_i(p_T)\left(M_{i,T}\, \delta T + M_{i,B}\, \delta\beta\right)
\biggr\rangle \\[0.4em]
&\quad \times (A\, \delta T + B\, \delta\beta) \\[0.6em]
= \; &\frac{1}{N_i} \Bigl[
  A\, \sigma_T^2\, \bigl(N'_{i,T}(p_T) - f_i(p_T)\, M_{i,T}\bigr) \\
&\quad + B\, \sigma_\beta^2\, \bigl(N'_{i,\beta}(p_T) - f_i(p_T)\, M_{i,B}\bigr)
\Bigr]
\end{aligned}
\end{equation}
with \(\sigma_T^2\equiv\langle\delta T^2\rangle\) and \(\sigma_\beta^2\equiv\langle\delta\beta^2\rangle\).
The variance of \(p_T\) reads $\hat\sigma_{p_T} = \sqrt{A^2\,\sigma_T^2 + B^2\,\sigma_\beta^2}.$
Finally, using the definition of $v_0(\pt)$, the parameterized form for a given particle species is given by

\begin{equation}
\boxed{%
  \begin{split}
    v_{0,i}(p_T)
    &= \frac{A\,\sigma_T^2\bigl[N_{i,T}'(p_T)-M_{i,T} f_{i}(p_T)\bigr]}
             {N_{i}(p_T)\sqrt{A^2\sigma_T^2+B^2\sigma_\beta^2}}\\
    &\quad
    +\;\frac{B\,\sigma_\beta^2\bigl[N_{i,\beta}'(p_T)-M_{i,B} f_{i}(p_T)\bigr]}
             {N_{i}(p_T)\sqrt{A^2\sigma_T^2+B^2\sigma_\beta^2}}\,.
  \end{split}
}
\label{eq:v0_bw}
\end{equation}
where
\begin{align}
N_{i,T}'(p_T)
&=
\frac{-A_{i}\,2\pi\,p_T\,m_{i,T}}{T^2}
\int_{0}^{1} r\,dr \int_{0}^{2\pi} d\phi\; \Bigl[ \notag\\
&\quad
p_T\sinh\rho \cdot
I_{1}\!\left(\tfrac{p_T\sinh\rho}{T}\right)
\cdot
K_{1}\!\left(\tfrac{m_{i,T}\cosh\rho}{T}\right) \notag\\
&\quad
+\, m_{i,T}\cosh\rho \cdot
I_0\!\left(\tfrac{p_T\sinh\rho}{T}\right)
\cdot
K'_{1}\!\left(\tfrac{m_{i,T}\cosh\rho}{T}\right)
\Bigr],
\label{eq:NTp}
\\[1em]
N_{i,\beta}'(p_T)
&=
\frac{A_{i}\,2\pi\,p_T\,m_{i,T}}{T\,(1-\beta_{s}^2)}
\int_{0}^{1} r^{2}\,dr \int_{0}^{2\pi} d\phi\; \Bigl[ \notag\\
&\quad
 p_T\cosh\rho \cdot
I_{1}\!\left(\tfrac{p_T\sinh\rho}{T}\right)
\cdot
K_{1}\!\left(\tfrac{m_{i,T}\cosh\rho}{T}\right) \notag\\
&\quad
+\, m_{i,T}\sinh\rho \cdot
I_0\!\left(\tfrac{p_T\sinh\rho}{T}\right)
\cdot
K'_{1}\!\left(\tfrac{m_{i,T}\cosh\rho}{T}\right)
\Bigr].
\label{eq:Nbp}
\end{align}
Note that in the absence of kinetic freeze-out temperature and radial expansion velocity fluctuations, i.e., \(\sigma_T = 0\) and \(\sigma_\beta = 0\), the observable $v_0(\pt)$ vanishes.

\begin{figure}[htbp]
  \centering
  \includegraphics[width=\linewidth]{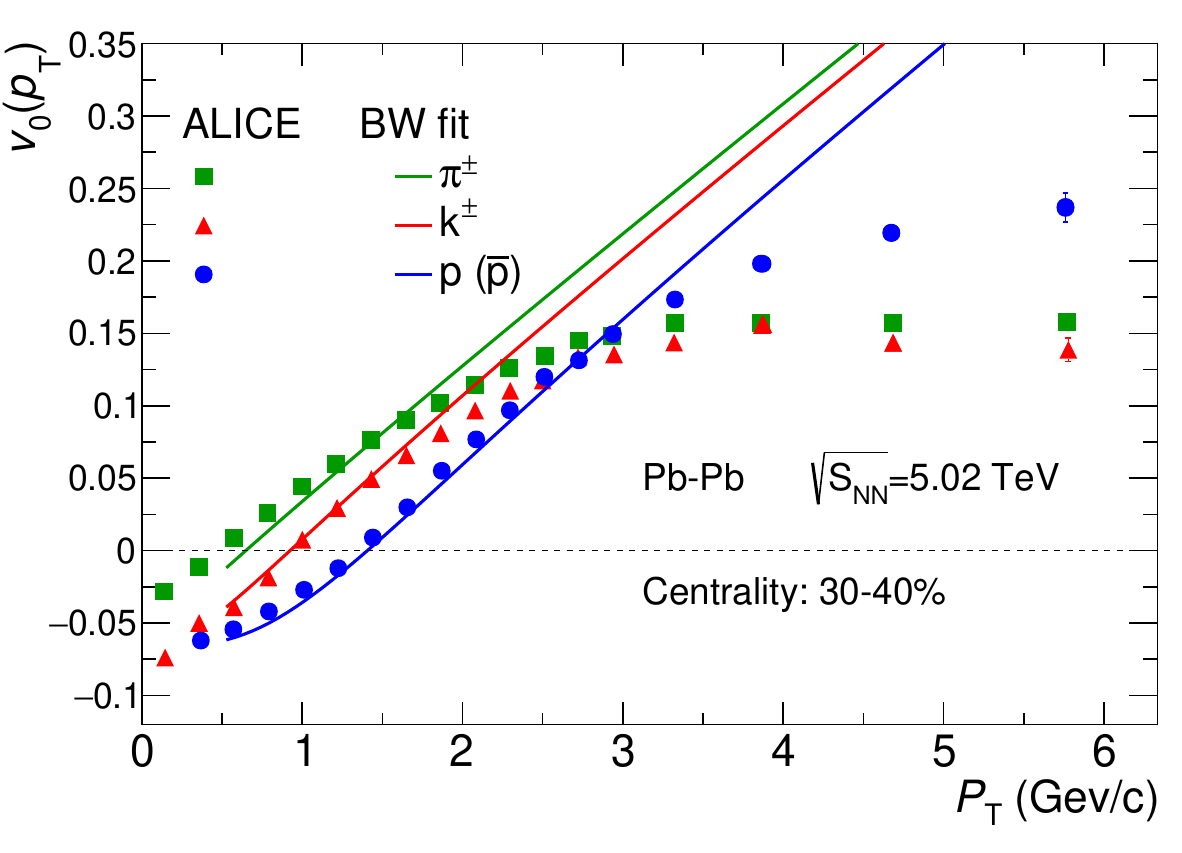}
  \caption{ \(v_{0}(\pt)\) fit to ALICE data in 30-40\% Pb–Pb collisions at \snn = 5.02 TeV according to Eq.~(\ref{eq:v0_bw}).}
  \label{fig:v0_fit}
\end{figure}

Based on Eq.~(\ref{eq:v0_bw}), we perform a simultaneous fit to the ALICE $v_0(\pt)$ data for $\pi$, $K$, and $p$ to extract the fluctuation amplitudes \(\sigma_T\) and \(\sigma_\beta\). In this fit, the parameters listed in Tab.~\ref{tab:params} are fixed, with the fitting performed over the \pt ranges [0.5, 1.5] GeV/$c$ for $\pi$, [0.5, 2.0] GeV/$c$ for $K$, and [0.5, 2.5] GeV/$c$ for $p$.
The fluctuation amplitudes are found to be
\begin{align*}
\sigma_T &= 0.0012 \pm 0.0006, \\
\sigma_\beta &= 0.0098 \pm 0.0021.
\end{align*}
These extracted fluctuations amount to approximately 1.1-1.2\% of the mean parameter values listed in Tab.~\ref{tab:params}.
 
\subsection{Mass ordering of $v_0(\pt)$}
\begin{figure}[htbp]
  \centering
  \includegraphics[width=\linewidth]{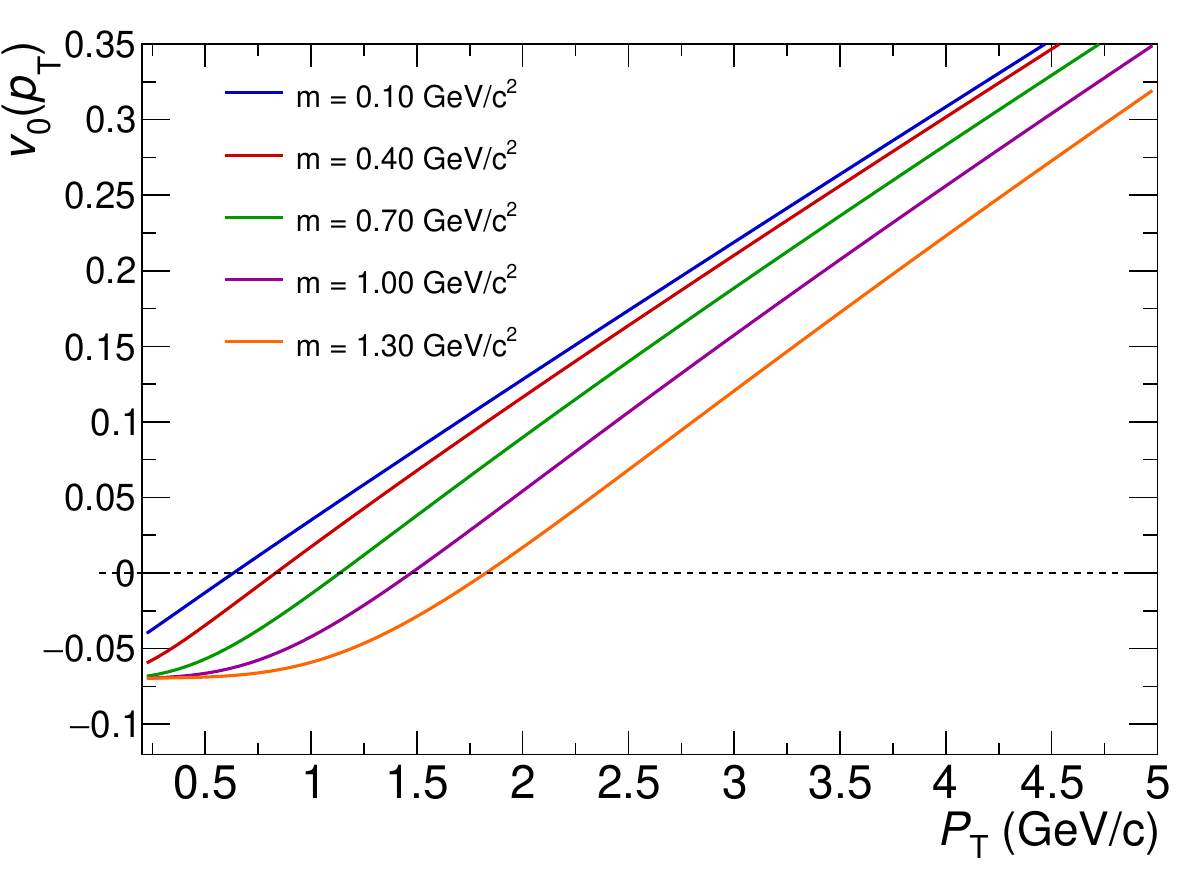}
  \caption{Systematic mass dependence in the BW model with \(\sigma_{\beta} = 0.0098\) and \(\sigma_T = 0.0012\), showing clear mass ordering across low- and intermediate-\(\pt\) range.}
  \label{fig:mass_ordering}
\end{figure}

The derived $v_0(\pt)$ from Eq.~(\ref{eq:v0_bw}) successfully reproduces the experimental trends and naturally exhibits a clear mass ordering in the low-\(p_T\) region, as shown in Fig.~\ref{fig:v0_fit}. Lighter particles yield larger $v_0$ values at a given \pt. To further validate this behavior, we systematically vary the particle mass and confirm that the ordering persists into the intermediate-\(p_T\) region, as illustrated in Fig.~\ref{fig:mass_ordering}. This mass-dependent structure parallels the well-known pattern observed in \(v_2(\pt)\), though the underlying mechanisms are different:
the mass ordering in \(v_{2}(\pt)\) arises from the mass-dependent response to radial expansion~\cite{mass_ordering_v2}. For a given radial expansion velocity, heavier particles are more effectively boosted toward higher \(\pt\), leading to a reduction in their low \(\pt\) yield. Since \(\rho_{2}\) enhances the flow velocity of in‑plane particles, the in-plane yield at low \(\pt\) is further suppressed, resulting in a smaller smaller \(v_{2}(\pt)\).

The mass dependence of $v_0(\pt)$, however, originates from fluctuations in the spectral shape. In the calculation of $v_0(\pt)$ for a given species, both \(\delta p_T\) and \(\sigma_{p_T}\) are evaluated over the full ensemble of particles. As per above derivation, the mass dependence arises primarily through the spectral fluctuation term
\begin{align}
v_0(\pt) 
&\propto \frac{\delta f(p_T)}{f(p_T)} 
= \delta \ln f(p_T) \notag = \delta \ln N(p_T) - \delta \ln N.
\end{align}
In the picture where fluctuations in radial expansion velocity drive $v_0(\pt)$, the multiplicity fluctuation term \(\delta\ln{N}\) contributes only marginally. Holding the variables \(r\),\(\rho\) and \(T\) fixed, we find
\begin{align}
\delta \ln f(p_T) \propto 
&\left(
  \frac{I_{1}\left( \frac{p_T \sinh \rho}{T} \right)}{I_{0}\left( \frac{p_T \sinh \rho}{T} \right)}
  \cdot \frac{p_T \cosh \rho}{T} \right. \notag\\
&\left. \quad
  +\,
  \frac{K_{1}'\left( \frac{m_T \cosh \rho}{T} \right)}{K_{1}\left( \frac{m_T \cosh \rho}{T} \right)}
  \cdot \frac{m_T \sinh \rho}{T}
\right)
\cdot \delta \rho,
\end{align}
where the second term decreases with increasing particle mass. As a result, the spectral response \(\delta \ln f(p_T)\) is suppressed for heavier particles, leading to a reduction in $v_0(\pt)$.

\begin{figure}[htbp]
  \centering
  \includegraphics[width=\linewidth]{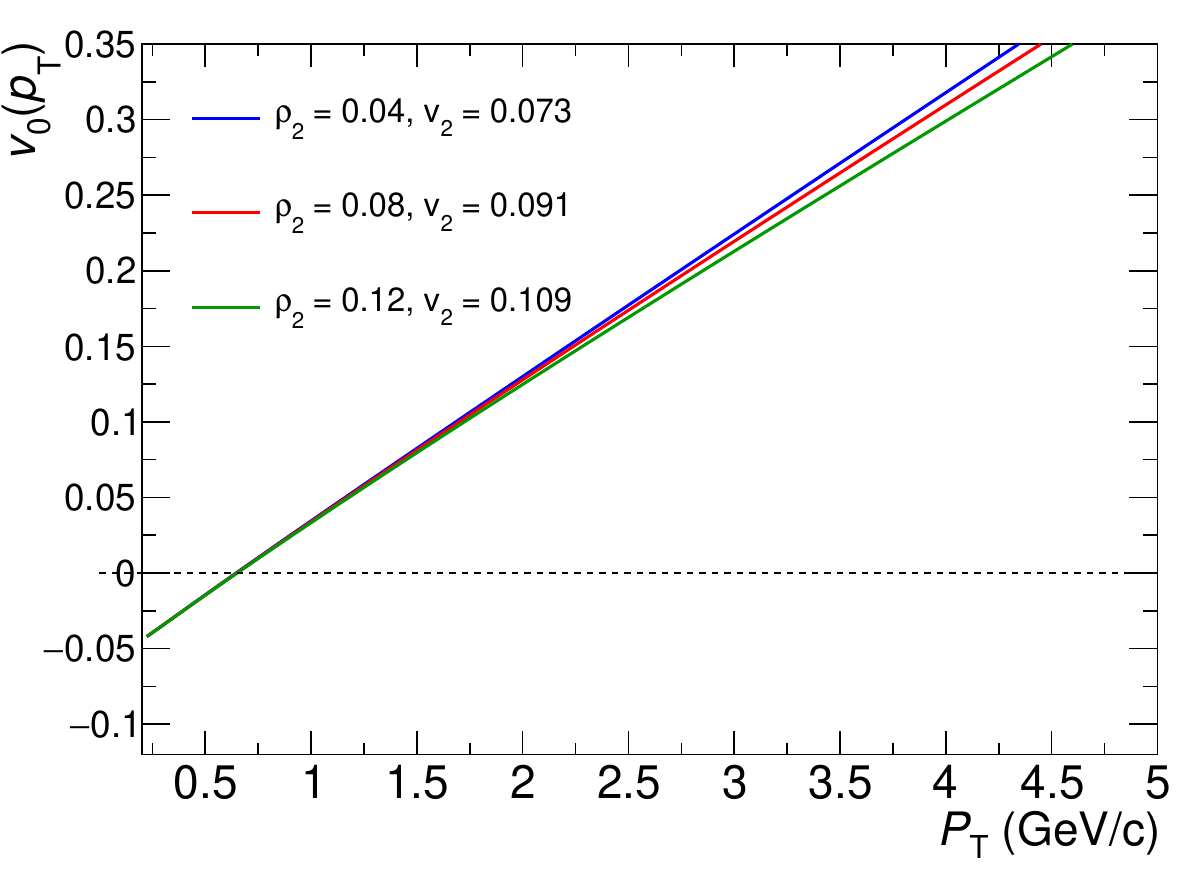}
  \caption{Systematic \(\rho_{2}\) variation in the BW model with \(\sigma_{\beta} = 0.0098\) and \(\sigma_T = 0.0012\), showing minimal change across low- and intermediate-\pt range.}
  \label{fig:rho2_dep}
\end{figure}

To further demonstrate that the mass ordering observed in $v_2(\pt)$ and $v_0(\pt)$ originates from different physical mechanisms, we perform a controlled study by varying \(\rho_2\), the parameter most directly responsible for azimuthal anisotropy, while keeping all other parameters fixed. The particle mass is set to the pion mass for simplicity. As shown in Fig.~\ref{fig:rho2_dep}, the resulting$v_0(\pt)$ exhibits almost no change in the low- to intermediate-\(p_T\) region (\(<2\)~GeV/$c$) across a wide range of \(\rho_2\), confirming that the mass ordering in $v_0(\pt)$ is an intrinsic feature of the spectral shape fluctuations rather than a consequence of azimuthal flow modulation. At \(p_T \approx 3\)~GeV/\(c\), a 50\% variation in \(v_2\) induces less than a 3\% change in $v_0(\pt)$, reinforcing the decoupling between two observables.

Meanwhile, we notice that in the intermediate- to high-\pt region, the BW model predicts nearly parallel $v_0(\pt)$ curves for different particle species, preserving the mass ordering. This behavior indicates that thermal freeze-out and collective expansion alone cannot explain the baryon–meson splitting observed at intermediate \pt. Additional mechanisms are therefore required; this will be discussed in the next section.

\section{$v_0(\pt)$ from quark transport and coalescence}

\begin{figure}[htbp]
  \centering
  \includegraphics[width=\linewidth]{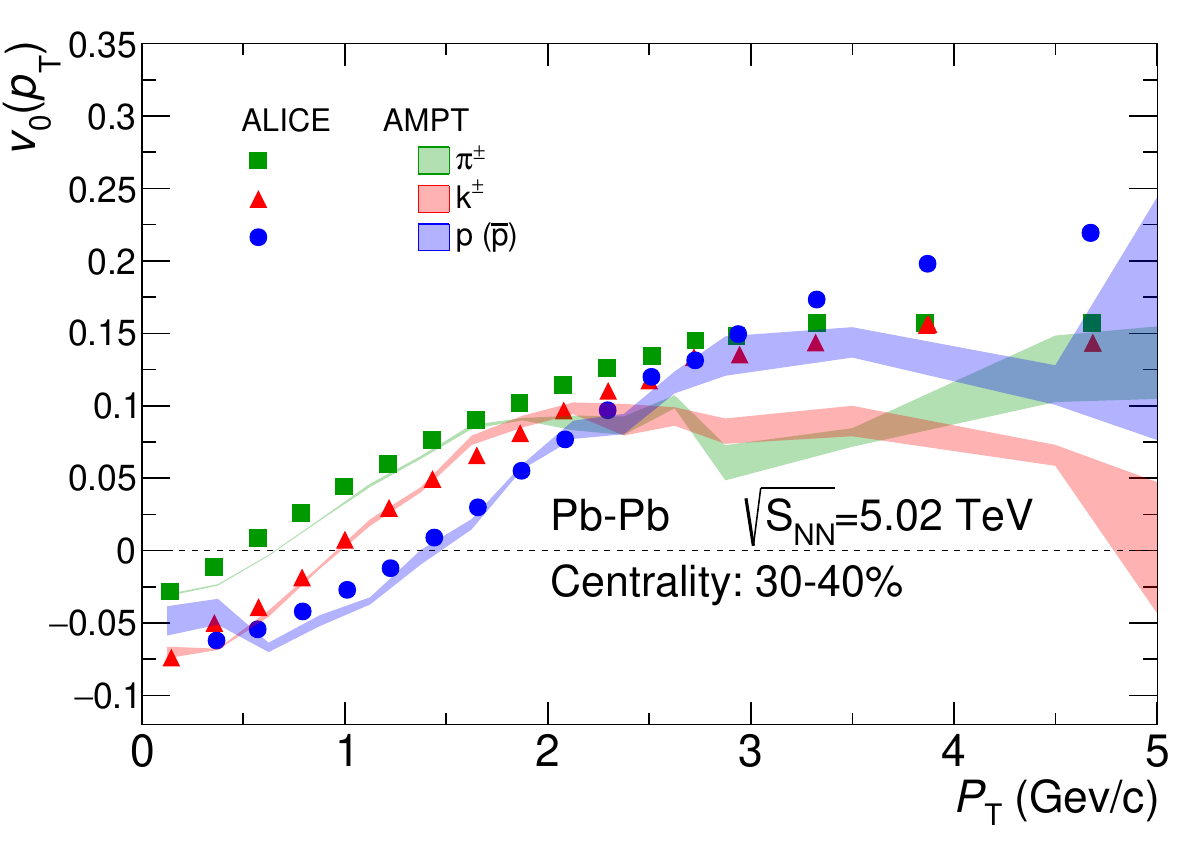}
  
  \caption{AMPT results for \(\pi\), \(K\), and \(p\) \(v_{0}(\pt)\), exhibiting clear mass ordering and baryon-meson splitting.}
  \label{fig:ampt_v0}
\end{figure}

To understand the intermediate-\(p_T\) behavior, the hybrid transport model AMPT~\cite{ampt1, ampt2} is employed. It is known for effectively describing the production and collective behavior of final-state hadrons, along with their dynamical evolution. The string melting version comprises four subroutines simulating successive stages of the collision: HIJING for initial parton condition, ZPC for partonic evolution, a simple quark coalescence for hadronization~\cite{coal}, and ART for hadronic rescatterings and interactions. 
We perform AMPT simulations of Pb--Pb collisions at \snn = 5.02 TeV in 30--40\% centrality with parameters tuned to reproduce the measured hadronic spectra and flow at LHC energies~\cite{ampt_par1, ampt_par2}. 
To suppress nonflow contributions and autocorrelations, we compute $v_0(\pt)$ with a pseudorapidity-gap of 0.3, consistent with the ALICE analysis~\cite{ALICE_v0,v0_2024}.
The resulting $v_0(\pt)$ for $\pi$, $K$ and $p$ is shown in Fig.~\ref{fig:ampt_v0}, compared with ALICE data. At low \pt, the simulation reproduces a clear mass ordering, which aligns with thermal expectations. Above 2.5 GeV/$c$, the $\pi$ and $K$ curves gradually converge, while the proton $v_0(\pt)$ continues to rise and separates upward. This baryon-meson splitting in the intermediate-\(p_T\) region is not captured by purely thermal models, highlighting the essential role of partonic dynamics and hadronizations. 

\begin{figure}[htbp]
  \centering
  \includegraphics[width=\linewidth]{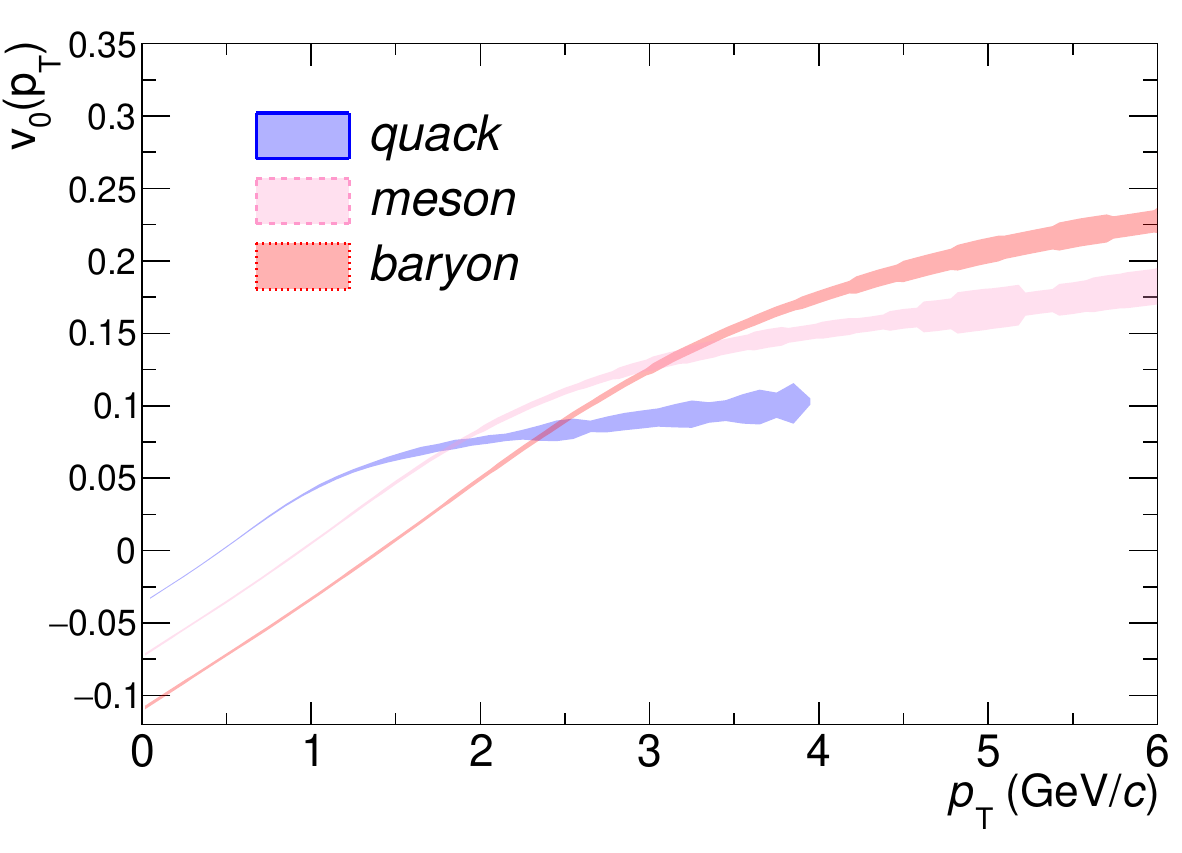}
  \caption{Meson and baryon \(v_{0}(\pt)\) calculated with Eq.~(\ref{eq:coalescence_v0}), compared with parton \(v_{0}(\pt)\) after ZPC stage.}
 \label{fig:parton_v0}
\end{figure}

Fig.~\ref{fig:parton_v0} shows the quark-level $v_0(\pt)$ before and after partonic scatterings. Before any interactions, partons exhibit a weak but non-zero $v_0(\pt)$, consistent with direct HIJING simulations~\cite{ALICE_v0}. After partonic scatterings and collective expansion, $v_0(\pt)$ is strongly enhanced, particularly in 0 -- 1.5 GeV/$c$. These partonic processes establish the basis for hadronic $v_0(\pt)$.

\begin{figure}[htbp]
  \centering
  \includegraphics[width=\linewidth]{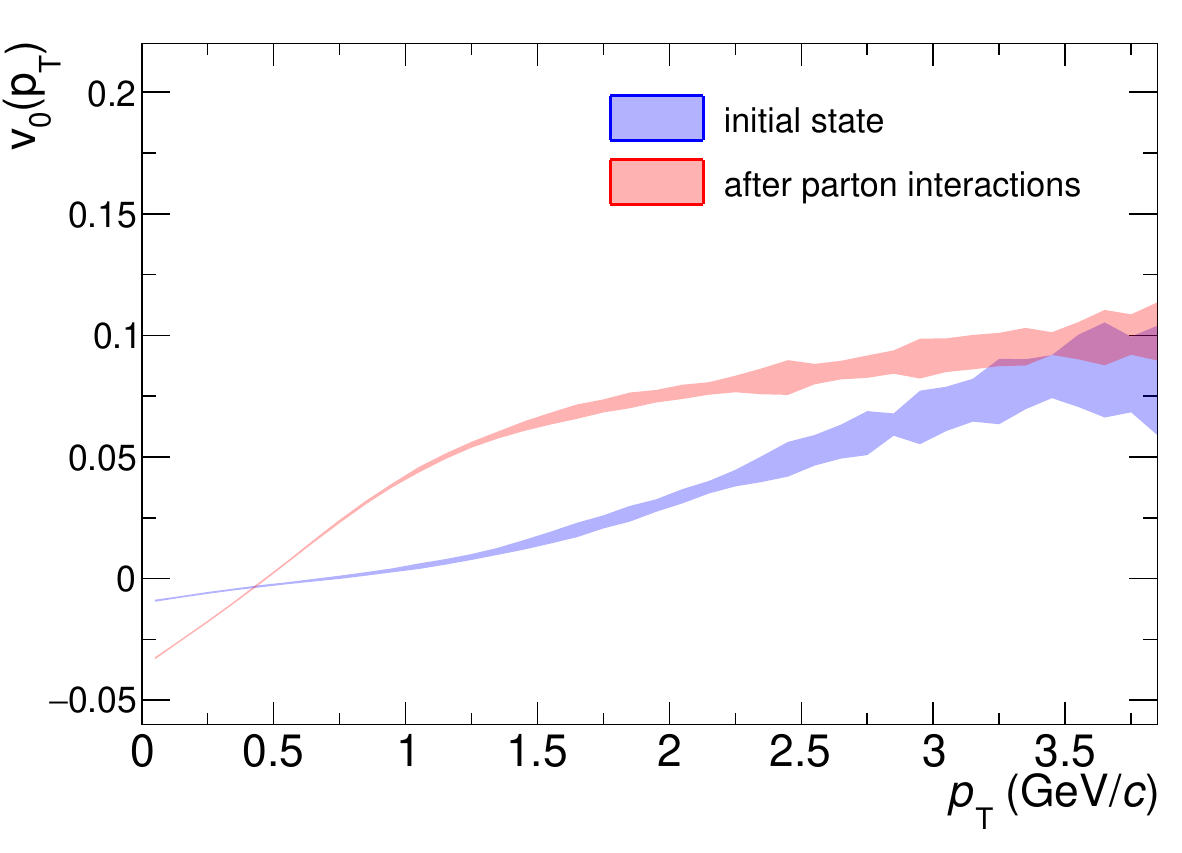}
  \caption{\(v_{0}(\pt)\) for light quark ($u, d$) in initial stage and after partonic interactions in 30--40\% Pb-Pb AMPT simulations.}
  \label{fig:ampt_v0t0}
\end{figure}

In a basic quark coalescence scenario~\cite{coal, coal2}, the invariant spectrum of produced hadrons is proportional to the product of the invariant spectra of its constituent quarks. For mesons (M) and baryons (B), the spectra are given by~\cite{Molnar:2003}:
\begin{equation}
\frac{dN_{M(B)}}{d^2\pt}(\vec{p}_{T}) = C_{M(B)}(\pt) \left[ \frac{dN_q}{d^2\pt}(\vec{p}_{T}/{2(3)}) \right]^{2(3)}
\label{eq:spec}
\end{equation}
where the coefficient $C_{M(B)}$ is the probability of forming meson (baryon). 
These coefficients are allowed to depend on $\pt$ in order to incorporate kinematic effects and the influence of strong radial flow~\cite{v2_in-high-pt}.
The fluctuations in the quark spectra can be written as
\begin{equation}
    \delta N_q(p_T,\beta) = N_{q,\beta}'(p_T,\beta)\delta \beta.
    \label{eq:quark_fluctuation}
\end{equation}
Assuming that fluctuations in the quark spectra are dominated by radial expansion velocity fluctuations, with multiplicity fluctuations neglected, the resulting fluctuations in the meson and baryon spectra read:
\begin{equation}
\delta N_M(p_T,\beta) = 2C_{M}(pt)N_{q}(\frac{\pt}{2})N_{q,\beta}'(\frac{\pt}{2},\beta)\delta \beta,
\label{eq:meson_fluct} \\
\end{equation}
\begin{equation}
\delta N_B(p_T,\beta) = 3C_{B}(pt)N_{q}^{2}(\frac{\pt}{3})N_{q,\beta}'(\frac{\pt}{3},\beta)\delta \beta,
\label{eq:baryon_fluct} \\
\end{equation}
\begin{equation}
\begin{aligned}
\delta \pt
&= \frac{1}{N_B+N_M} \int \! \mathrm{d}p_T\, p_T \left[ \delta N_B(p_T,\beta) + \delta N_M(p_T,\beta) \right] \\
&= \frac{1}{N_B + N_M} \int \! \mathrm{d}p_T\, p_T \left[
  3C_{B}(p_T)N_{q}^{2}\left(\frac{\pt}{3}\right)N_{q,\beta}'\left(\frac{\pt}{3},\beta\right) \right. \\
&\qquad\qquad\qquad\left. +\, 2C_{M}(p_T)N_{q}\left(\frac{\pt}{2}\right)N_{q,\beta}'\left(\frac{\pt}{2},\beta\right)
\right] \delta\beta. \\
\end{aligned}
\end{equation}
According to the definition in Eq.~(\ref{eq:v0_original}), we have
\begin{equation}
\boxed{
  v_{0}^{h}(\pt^{h}) = n_{q} \; v_{0}^{q}\!\Bigl(\frac{\pt^{h}}{n_{q}}\Bigr),
  }
  \label{eq:coalescence_v0}
\end{equation}
with $n_{q}=2$ for mesons and $n_{q}=3$ for baryons.
Note that Eq.~\eqref{eq:coalescence_v0} is structurally identical to the well-established relation for elliptic flow, $v_2^h(p_T^h)=n_q\,v_2^q(p_T^h/n_q)$, which has been widely used to interpret the baryon-meson splitting observed for $v_2(\pt)$.

Based on this expression and using the \(u/d\) quark $v_0(\pt)$ extracted from AMPT after partonic scatterings as input, we calculate the $v_0(\pt)$ for mesons and baryons, shown in Fig.~\ref{fig:parton_v0}. Both mass ordering and baryon-meson splitting are clearly obtained. Notably, this calculation yields a crossing point near \(3\,\mathrm{GeV}/c\), which closely aligns with experimental data. Minor differences between this direct calculation and the full AMPT results shown in Fig.~\ref{fig:ampt_v0} can be attributed to the spatial-proximity criterion governing quark coalescence in AMPT and the more realistic treatment of hadronic evolution.

\section{Summary}

In this letter, we investigate the observable $v_0(\pt)$ for identified particles, recently measured by ALICE in Pb--Pb collisions at \snn = 5.02 TeV, using a blast-wave model and AMPT model.
Within the BW framework, we introduce event-by-event fluctuations of the freeze-out temperature $T$ and the radial flow velocity $\rho_{0}$ and derive an analytic expression for $v_0(\pt)$.
A simultaneous fit to ALICE $\pi$, $K$, and $p$ measurements reproduces the observed low-\pt mass ordering, which we demonstrate originates from the thermal spectral response to radial flow fluctuations rather than azimuthal anisotropy.
Using AMPT, we qualitatively reproduce the intermediate-\pt baryon-meson splitting and trace its origin to quark dynamics. Partonic scatterings enhance $v_0(\pt)$, and a generalized coalescence mechanism can naturally explain the splitting.

In summary, $v_0(\pt)$ serve as a novel observable offering an independent cross-check for partonic and hadronic collectivity. This work provides an essential piece for a deeper understanding of $v_0(\pt)$. Future work may extend to broader collective phenomena, such as NCQ scaling and event shape engineering, which have been extensively studied in $v_2$.

\section*{Acknowledgements}
We are grateful to Swati Saha and other colleagues in the ALICE Collaboration for inspiring this work. We thank Guangyou Qin for enlightening discussions and Chen Zhong for his dedicated maintenance of the computing resources. This work is supported by the National Natural Science Foundation of China (Nos. 12322508, 12061141008, 12147101), the National Key Research and Development Program of China (No. 2024YFA1610802), and the Science and Technology Commission of Shanghai Municipality (No. 23590780100).

\appendix


\end{document}